\documentclass[useAMS,usenatbib]{mn2e}
\newcommand{\be}{\begin{equation}}
\newcommand{\ee}{\end{equation}}
\newcommand{\ba}{\begin{eqnarray}}
\newcommand{\ea}{\end{eqnarray}}

\newcommand{\watertrans}{\hbox{H$_{\rm 2}$O$~(2_{\rm 02}-1_{\rm 11}$)}}

\def\simless{\mathbin{\lower 2.5pt\hbox
   {$\rlap{\raise 4.5pt\hbox{$\char'074$}}\mathchar"7218$}}}
\def\simgrt{\mathbin{\lower 2.5pt\hbox
   {$\rlap{\raise 4.5pt\hbox{$\char'076$}}\mathchar"7218$}}}
\include{epsf}

\title[Reconstruction of H-ATLAS J090311.6$+$003906 using ALMA]
{Revealing the complex nature of the strong gravitationally lensed 
system H-ATLAS J090311.6$+$003906 using ALMA}
\author[S. Dye et al.]
{\parbox{\textwidth}{S. Dye$^{1}$\thanks{E-mail: simon.dye@nottingham.ac.uk},
C. Furlanetto$^{1,2}$, 
A. M. Swinbank$^{3}$, 
C. Vlahakis$^{4,5}$,
J. W. Nightingale$^{1}$,
L. Dunne$^{6,7}$, 
S. A. Eales$^{8}$, 
Ian Smail$^{3}$, 
I. Oteo$^{7,9}$, 
T. Hunter$^{10}$, 
M. Negrello$^{11}$,  
H. Dannerbauer$^{12}$, 
R. J. Ivison$^{7,9}$,
R. Gavazzi$^{13}$, 
A. Cooray$^{14}$,
P. van der Werf$^{15}$}
\vspace{4mm}\\
$^{1}$School of Physics and Astronomy, Nottingham University,
University Park, Nottingham, NG7 2RD, UK\\
$^{2}$CAPES Foundation, Ministry of Education of Brazil, Bras\'ilia/DF, 
70040-020, Brazil\\
$^{3}$Centre for Extragalactic Astronomy, Durham University,
South Road, Durham DH1 3LE, UK \\
$^{4}$Joint ALMA Observatory, Alonso de C\'{o}rdova 3107, 
Vitacura, Santiago, Chile\\
$^{5}$European Southern Observatory, Alonso de C\'{o}rdova 3107, 
Vitacura, Santiago, Chile\\
$^{6}$Department of Physics and Astronomy, University of Canterbury,
Private Bag 4800, Christchurch, 8140, New Zealand\\
$^{7}$Institute for Astronomy, Royal Observatory Edinburgh, 
Blackford Hill, Edinburgh, EH9 3HJ, UK.\\
$^{8}$School of Physics and Astronomy, Cardiff University,
Queen's Buildings, The Parade, Cardiff, CF24 3AA, UK\\
$^{9}$European Southern Observatory, Karl-Schwarzschild-Str. 2, 
Garching, Germany\\
$^{10}$National Radio Astronomy Observatory, 520 Edgemont Rd, 
Charlottesville, VA, 22903, USA\\
$^{11}$INAF, Osservatorio Astronomico di Padova, Vicolo Osservatorio 5,
I-35122 Padova, Italy\\
$^{12}$Universit\"at Wien, Institut f\"ur Astrophysik,
T\"{u}rkenschanzstrasse 17, 1180 Wien, Austria\\
$^{13}$Institut d'Astrophysique de Paris, UMR7095 CNRS-Universit\'{e} 
Pierre et Marie Curie, 98bis bd Arago, F-75014 Paris, France\\
$^{14}$Astronomy Department, California Institute of Technology,
MC 249-17, 1200 East California Boulevard, Pasadena, CA
91125, USA\\
$^{15}$Leiden Observatory, Leiden University, P.O. Box 9513, NL-2300 RA Leiden, The Netherlands
}

\begin{document}

\date{}

\pagerange{\pageref{firstpage}--\pageref{lastpage}} 
\pubyear{2014}

\maketitle

\label{firstpage}

\begin{abstract}
We have modelled Atacama Large Millimeter/sub-millimeter Array (ALMA)
long baseline imaging of the strong gravitational lens system H-ATLAS
J090311.6$+$003906 (SDP.81). We have reconstructed the distribution of
band 6 and 7 continuum emission in the $z=3.042$ source and we have
determined its kinematic properties by reconstructing CO(5-4) and
CO(8-7) line emission in bands 4 and 6.  The continuum imaging reveals
a highly non-uniform distribution of dust with clumps on scales of
$\sim 200$\,pc. In contrast, the CO line emission shows a relatively
smooth, disk-like velocity field which is well fit by a rotating disk
model with an inclination angle of $(40 \pm 5)^\circ$ and an
asymptotic rotation velocity of 320\,kms$^{-1}$.  The inferred
dynamical mass within 1.5\,kpc is
$(3.5\pm0.5)\times10^{10}$\,M$_\odot$ which is comparable to the total
molecular gas masses of $(2.7\pm0.5)\times10^{10}$\,M$_\odot$ and
$(3.5\pm0.6)\times10^{10}$\,M$_\odot$ from the dust continuum emission
and CO emission respectively.  Our new reconstruction of the lensed
HST near-infrared emission shows two objects which appear to be
interacting, with the rotating disk of gas and dust revealed by ALMA
distinctly offset from the near-infrared emission.  The clumpy nature
of the dust and a low value of the Toomre parameter of $Q \sim 0.3$
suggest that the disk is in a state of collapse. We
estimate a star formation rate in the disk of
$470\pm80$\,M$_\odot$/yr with an efficiency $\sim 65$ times greater
than typical low-redshift galaxies. Our findings add to the growing
body of evidence that the most infra-red luminous, dust obscured
galaxies in the high redshift Universe represent a population of
merger induced starbursts.

\end{abstract}


\section{Introduction}

Our understanding of high redshift sub-millimetre (submm) bright
galaxies (SMGs) has grown immensely since their discovery nearly two
decades ago \citep{smail97,hughes98,barger98}. The fact that
approximately half of the total energy output from stars within the
observable history of the Universe has been absorbed by dust and
re-emitted at submm wavelengths \citep{puget96,fixsen98} and that SMGs
represent the most active sites of dusty star formation at high
redshifts, indicates that their role in early galaxy formation is an
important one.

Morphological and kinematical measurements of SMGs have led many
studies to conclude that they are a more energetic version of more
local ultra-luminous infrared galaxies \citep[ULIRGs; e.g.][]{engel10,
  swinbank10,alaghband12,rowlands14}. However, observations have been
limited by the low imaging resolution typically offered by submm
facilities, forcing detailed investigations to turn to other
wavelengths which permit higher resolution.  With strong attenuation
over ultraviolet to near-infrared wavelengths, where imaging technology
has benefited from a longer period of development, this has proven
challenging. Hence, a reliance has traditionally been made on
correlations with other wavelengths which often only result in
indirect diagnostics of the internal energetics of the physical
processes at work in these galaxies.

A powerful diagnostic which gives unique insight into the
star formation processes in galaxies in general is measurement of
molecular gas \citep[see for example][and references
therein]{papadopoulos14}. In particular, in the more extreme
environments of ULIRG and SMG interiors where strong feedback from
star-formation and shock-heating of molecular gas by supernovae are
dominant processes, star formation models can be put through the most
rigorous of tests \citep[e.g.][]{papadopoulos11}. In this way,
examination of the quantity and kinematical properties of molecular
gas provides a direct probe of the mode of star formation. Following
this approach, some studies have concluded that star formation
mechanisms in early systems are distinctly different from those seen
in more local systems \citep[e.g.][]{bournaud09,jones10}.

A detailed study to measure the properties of star formation in
distant SMGs requires two key ingredients.  Firstly, observations must
be carried out at submm wavelengths, where most of the bolometric
luminosity is emitted, to allow emission from the dust enshrouded
molecular gas to be detected. Secondly, the observations must be of
sufficient angular resolution to isolate the dynamics of the $\sim
200$\,pc gravitationally unstable regions usually found in their
star-forming disks \citep{downes98}.

To provide a sample of suitable SMGs for such investigation, the large
area surveys carried out recently by the Herschel Space Observatory,
such as the Herschel Astrophysical Terahertz Large Area Survey
\citep[H-ATLAS;][]{eales10} and the Herschel Extragalactic
Multi-tiered Extragalactic Survey \citep{oliver12} along with the
survey at millimetre wavelengths carried out at the South Pole
Telescope \citep{carlstrom11,vieira13} now provide a bountiful
supply of high redshift dusty star bursts for detailed study.
Obtaining the required high resolution imaging in the submm is
now made possible using the Atacama Large Millimetre/sub-millimetre
Array (ALMA).

A particularly compelling use of ALMA for these purposes is to target
strongly lensed SMGs. The submm has long been suspected to harbour a
rich seam of strongly lensed galaxies due to a high magnification bias
resulting from their steep number counts \citep{blain96,negrello07}.
Thanks to the aforementioned mm/submm surveys, such suspicions have
now been verified \citep{vieira10,negrello10,hezaveh11,wardlow13}.
The intrinsic flux and spatial magnifications by factors of $\sim 10 -
30$ inherent in strongly lensed systems therefore combine with the
use of ALMA to provide the highest possible resolution and
signal-to-noise imaging of SMGs currently achievable by a considerable
margin.

In this paper, we report our analysis of the recently released ALMA
science verification observations of the strong lens system H-ATLAS
J090311.6$+$003906 (SDP.81), one of the first five strongly lensed
submm sources detected in the H-ATLAS data \citep{negrello10}. The
system was subsequently followed up in the near-infrared using the
Hubble Space Telescope \citep[HST; see] [for details of these
  observations; N14 hereafter]{negrello14}.  Lens modelling of the
system by \citet[][D14 hereafter]{dye14} showed that the observed
Einstein ring can be explained by a single component of emission in
the source plane.  The purpose of this paper is to exploit the high
resolution ALMA imaging of the highly magnified lensed source to
determine its physical properties. One of our key questions is how the
rest-frame optical source emission reconstructed by D14 relates to the
reconstructed submm emission detected by ALMA.

The layout of this paper is as follows: Section \ref{sec_data}
outlines the data. In Section \ref{sec_modelling} we describe the
modelling procedure used to obtain the results which are given in
Section \ref{sec_results}. We discuss our findings in Section
\ref{sec_discussion} and summarise the major results of this work in
Section \ref{sec_summary}.  Throughout this paper, we assume the
following cosmological parameters; ${\rm H}_0=67\,{\rm
  km\,s}^{-1}\,{\rm Mpc}^{-1}$, $\Omega_m=0.32$,
$\Omega_{\Lambda}=0.68$ \citep{planck14}.

\section{Data}
\label{sec_data}

\subsection{ALMA data}

ALMA Science Verification data on SDP.81 were taken from the
ALMA Science Portal\footnote{http://www.almascience.org} (ASP).  We
give an overview of those data here, although more details can be
found in \citet{vlahakis15}.

SDP.81 was observed in October 2014 as part of ALMA's Long Baseline
Campaign, using between 22 and 36 12\,m-diameter antennas and ALMA's
band 4, 6 \& 7 receivers. The band 4 observations had the fewest total
number of antennas (a maximum of 27 compared to a maximum of 36 in the
other bands), although the 21-23 element long baseline configuration
was similar in all three bands. 

Four 1.875\,GHz bandwidth spectral windows were used, over a total
bandwidth of 7.5\,GHz. In each observing band, one or two spectral
windows covered a spectral line, with the remaining spectral windows
used for continuum. The band 4, 6 and 7 data include the redshifted
CO(5-4) ($v_{\rm rest}$ = 576.267\,GHz), CO(8-7) ($v_{\rm rest}$ =
921.799\,GHz), and CO(10-9) ($v_{\rm rest}$ = 1151.985\,GHz) lines,
respectively, as well as rest frame 250\,$\mu$m, 320\,$\mu$m and
500\,$\mu$m continuum. The band 6 data also include the redshifted
low-excitation water line \watertrans\/ ($v_{\rm rest}$ = 987.927\,GHz;
$E_{\rm up}$ = 101 K) but we leave analysis of this feature for
future work (see Section \ref{sec_summary}).

%Various quasars were used for bandpass, phase and flux calibration 
%\citep[see][for more details]{vlahakis15}. The phase centre was 09$^{
%h}$03$^{m}$11$^{s}$.61 $+$00${\degr}$39${\arcmin}$06${\arcsec}$.7 (J2000). 

The calibration and imaging of the data is described in the scripts
provided on the ASP. These were carried out
using the Common Astronomy Software Application package
\citep[{\tt CASA\footnote{
      http://casa.nrao.edu}};][]{mcmullin07}.  A {\it robust}=1
weighting of the visibilities was used. Line-free channels were used
to subtract the continuum emission from the CO data. The CO line data
were imaged using rest frequencies corresponding to $z=3.042$ and were
$uv$-tapered to a resolution of $\sim$170\,mas (1000\,k$\lambda$), since
the high-resolution CO data has relatively low signal-to noise. In the
case of {\hbox{H$_{\rm 2}$O}}, the data were $uv$-tapered to
200~k$\lambda$ (providing an angular resolution of $\sim$0.9\arcsec)
in order to achieve a reasonable detection. The resulting RMS
noise levels are 0.20\,mJy and 0.15\,mJy per 21\,kms$^{-1}$ channel 
for CO(5-4) and CO(8-7) respectively. 

We also carried out our own independent imaging of the calibrated
visibility data supplied via the ASP. We attempted a variety of
different tapers and cleaning parameters but found the ASP data to be
already optimal for our purposes. We note also that in this paper we
have used the band 4 image cube later staged on the ASP on March 2nd
2015 with correctly subtracted continuum.

For the purposes of our lens modelling, we binned the band 6 and
band 7 continuum images from a pixel scale of $0.005''$ to a pixel
scale of $0.01''$. This not only increases modelling efficiency, but
also lessens image pixel covariance (see Section \ref{sec_modelling}). In
the modelling, we assumed the synthesised beam sizes prescribed in the
ALMA data themselves; 155$\times$121\,mas and 169$\times$117\,mas for
CO(5-4) and CO(8-7), respectively.

Panels (a) and (d) in Figure \ref{band6+7_cont_recon} show the
ALMA band 6 and band 7 continuum images.

\subsection{Near-IR data}

We have re-analysed the Hubble Space Telescope (HST)
imaging\footnote{The HST imaging was acquired with the Wide Field
  Camera 3 (WFC3) in Cycle 18 under proposal 12194 (PI Negrello).} of
SDP.81 modelled by D14. We have applied our modelling to the
deeper F160W image which has a total exposure time of 4418\,s. The
image was reduced using the IRAF {\tt MultiDrizzle} package 
resulting in an image resampled to a pixel scale of $0.064''$.

We have post-processed the data in two different ways.  Firstly, we
carried out an independent removal of the lens galaxy light prior to
lens modelling using the {\tt GALFIT} software \citep{peng02}. In
doing so, we have revealed additional structure in the F160W data to
the south of the lens.  Since this influences our new interpretation
of the characteristics of the lensed source, we include this
additional structure in our lens modelling with a larger image plane
to encompass it (see section {\ref{sec_src_morph} for more details).

Secondly, we applied a small astrometric shift of $\sim 0.1''$ to
align to the ALMA data. We determined this shift using the Sloan
Digital Sky Survey data release 10 \citep{ahn14} to identify stars in
the region covered by the HST image and then tied the resulting
stellar catalogue to the two Micron All Sky Survey
\citep{skrutskie06}. We note that this astrometric alignment agrees
very precisely with the independent alignment determined by shifting
the best fit ALMA lens model centre to the centroid of the observed
F160W flux.

In the lens modelling of the F160W data, we used a point spread
function created by the {\tt TinyTim} software package
\citep{krist93}. Panel (a) of Figure \ref{hst_recon}
shows the reprocessed F160W data with the band 6 continuum overlaid
as contours.

\section{Lens Modelling}
\label{sec_modelling}

We carried out our lens modelling in the image plane rather than the
$uv$ plane. There are two main advantages to this approach. The first
is that the image can be masked to limit calculation of the goodness
of fit to those parts of the image where there is detected
emission from the lensed source.  This gives a considerably more
sensitive figure of merit for the fitting than working in the $uv$ plane
where modelling necessarily fits to visibilities that largely describe
extended areas of background sky. The second is of particular
relevance to the ALMA dataset under analysis in this paper; modelling
in the image plane is vastly more efficient than working directly with
the extremely large visibility dataset which has to be trimmed in
Fourier space anyway to ensure that the modelling process is feasible
\citep[e.g.][]{rybak15}.

The disadvantage of working in the image plane is that the pure
interferometric visibilities are not directly modelled, but their
modulated Fourier transform instead. A side effect of this is that
image pixels are correlated by the beam which biases image plane
modelling if the uncertainties do not take the covariance of image
pixels into account. However, in the case of the ALMA data modelled in
this paper, the beam size is comparable to the image pixel scale and
so this covariance is relatively low.  The covariance is lowered even
further by our use of the $2 \times 2$ pixel binned version of the
band 6 and band 7 science verification continuum image
data. Furthermore, the ALMA data have a very high coverage of the $uv$
plane which significantly reduces errors in the image plane.

Any bias resulting from ignoring covariance in the image plane will
affect the overall normalisation of the figure of merit. Whilst this
effect will be small in the current ALMA data, this will still prevent
a fair comparison between different lens model parameterisations in
principle. However, for a fixed parameterisation such as that used in
the present work, the relative difference in the figure of merit
between different sets of parameter values remains unaffected. In this
way, a reliable best fit lens model can still be found and hence image
plane covariance is not a concern in this regard.

\begin{figure*}
\epsfxsize=18cm
{\hfill
\epsfbox{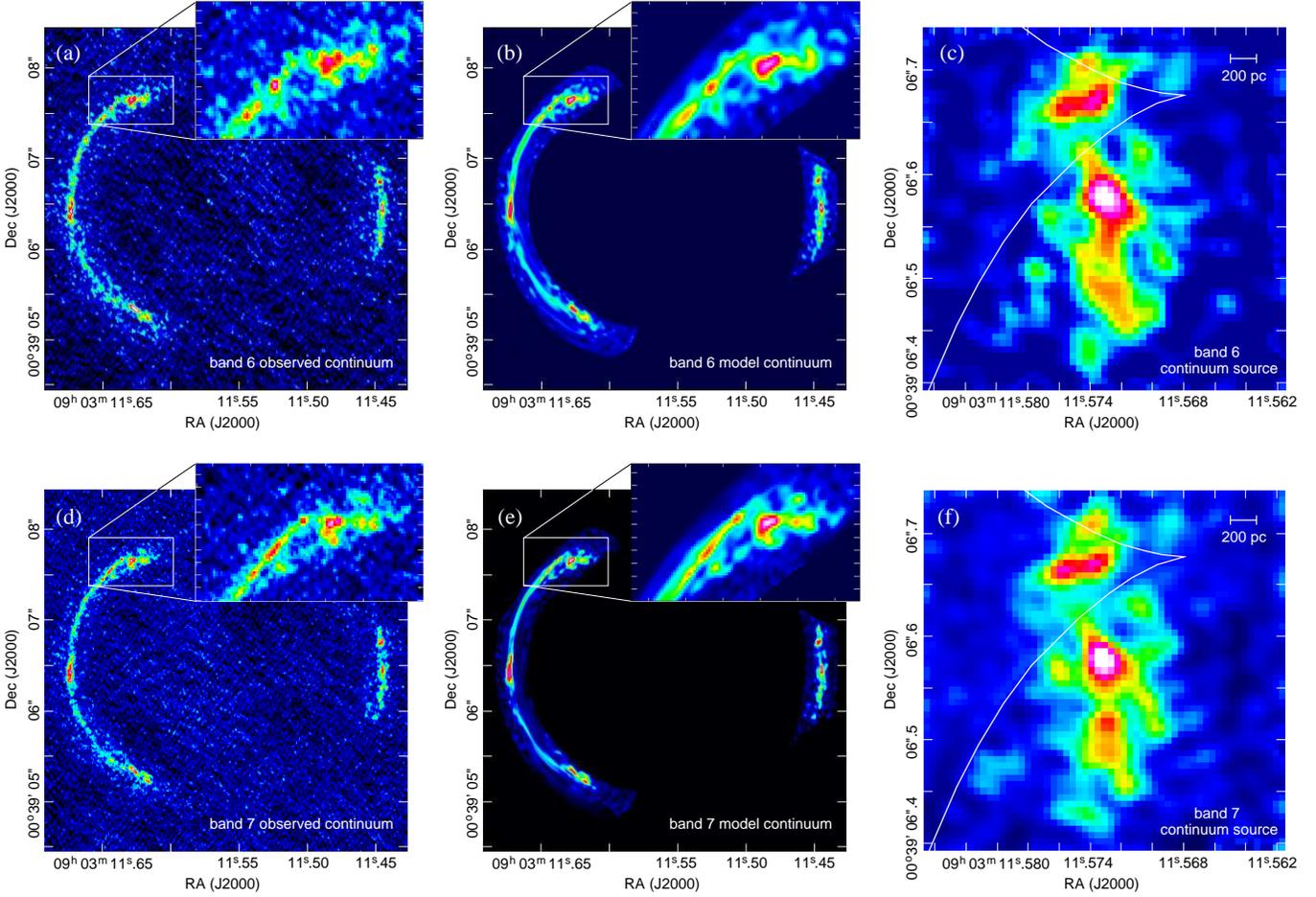}
\hfill}
\epsfverbosetrue
\caption{Panels (a) to (c) show the reconstruction of the band 6
  continuum image and panels (d) to (f) show the band 7 continuum
  reconstruction. The observed images are shown in panels (a) and (d),
  the model images of the reconstructed source are shown in panels (b)
  and (e) and the reconstructed sources are shown in panels (c) and
  (f). The white line in the source maps shows the position of the
  caustic. The inset zooms indicate the fidelity of the
  reconstruction.}
\label{band6+7_cont_recon}
\end{figure*}

We have verified that these assumptions are valid with the ALMA data
analysed in this paper using the following procedure.  Firstly, as we
discuss below, we located the best fit lens model by application of
the semi-linear method in the image plane to the $2 \times 2$ binned
version of the band 7 continuum image. We then transformed the best
fit model lensed image to the $uv$ plane by using {\tt MIRIAD uvmodel}
to produce a simulated visibility dataset for the same $uv$ coverage
as in the ALMA dataset. Using the ALMA visibilities, their
uncertainties and the model visibilities, we computed $\chi^2$. We
then varied different lens model parameters, stepping away from the set
which provide the best fit in the image plane, generating new images
each time and transforming to the $uv$ plane to measure how
$\chi^2$ varied.  We found that although there is a slight offset in
parameter space between the image plane minimum-$\chi^2$ and the $uv$
plane minimum-$\chi^2$, this is within the parameter uncertainties.

\subsection{Lens modelling procedure}

We used the latest implementation of the semi-linear inversion
method \citep{warren03} as described by \citet{nightingale15}.
The crux of the method is the manner in which the source plane
discretisation adapts to the magnification produced by a given
set of lens model parameters. By introducing a random element to
the discretisation and by ensuring that the discretisation maps
exclusively back to only those areas in the image within the mask,
the method eradicates biases in lens model parameter estimation.
Crucially, the method removes a significant bias in the inferred value
of the logarithmic slope of power-law mass density profiles which occurs
when semi-linear inversion is used with a fixed source plane size
and/or a fixed source plane pixellisation 
\citep[see][for more details]{nightingale15}.

We have also used simultaneous reconstruction of multiple source
planes from multiple images as described in D14.  We
attempted a variety of different combinations of data, but found that
there were no clear improvements on lens model constraints beyond
using a dual reconstruction of the ALMA band 6 and band 7 continuum image
data.

For our lens model, we adopted a single smooth power-law profile with
a volume mass density of the form $\rho \propto r^{-\alpha}$. In
utilising this profile, it is assumed that the power-law slope,
$\alpha$, is scale invariant. This assumption appears to be reasonable
on the scales probed by strong lensing as demonstrated by a lack of
any trend in slope with the ratio of Einstein radius to effective
radius \citep{koopmans06,ruff11}.

\begin{figure*}
\epsfxsize=18cm
{\hfill
\epsfbox{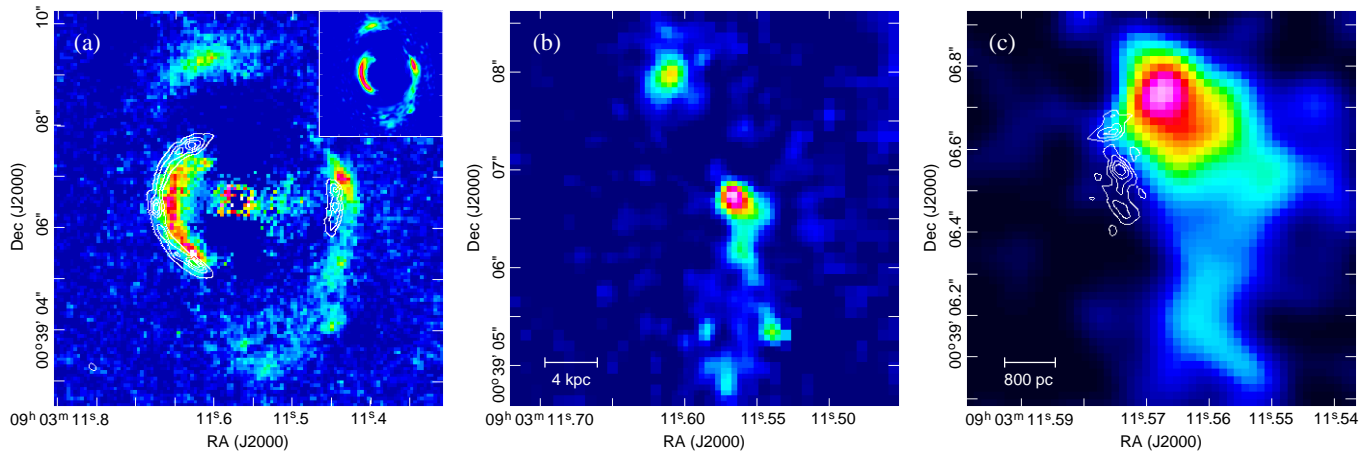}
\hfill}
\epsfverbosetrue
\caption{Source reconstruction of the HST F160W data. Panel (a) shows
  the observed image (note that the residual noise left from lens
  galaxy light subtraction is masked out in the lens modelling) with
  band 6 continuum contours overlaid and the inset panel shows the
  lensed image of the reconstructed source. In panel (b) we show the
  full scale of the reconstructed source that fits the tidal
  debris-like emission seen in the observed image. Finally, panel (c)
  shows a zoomed-in section of the source overlaid with band 6
  continuum source contours.}
\label{hst_recon}
\end{figure*}

The corresponding projected mass density profile used to calculate
lens deflection angles is therefore the
elliptical power-law profile introduced by
\citet{Ka93} with a surface mass density, $\kappa$, given by
\be
\kappa=\kappa_0\,({\tilde r}/{\rm 1kpc})^{1-\alpha} \, .
\ee
Here, $\kappa_0$ is the normalisation surface mass density (the
special case of $\alpha=2$ corresponds to the singular isothermal
ellipsoid, SIE). The radius ${\tilde r}$ is the elliptical radius defined
by ${\tilde r}^2 =x^{\prime2}+y^{\prime2}/\epsilon^2$ where $\epsilon$
is the lens elongation defined as the ratio of the semi-major to
semi-minor axes. Three further parameters define the lens mass profile:
the orientation of the semi-major axis measured in a
counter-clockwise sense from north, $\theta$, and the coordinates of
the centre of the lens in the image plane, $(x_c,y_c)$. Finally, following
the findings of D14, we include an external shear component
which is described by the shear strength $\gamma$ and orientation
$\theta_\gamma$ measured counter-clockwise from north.

As this paper is concerned primarily with the properties of the lensed
source, we have not considered more complicated lens models. For
example, one possibility is to model the lens using a cored density
profile. Whilst there is continuum emission detected at the centre of
the ring where a core would be expected to produce an image, our
measurements of the spectral index indicate that this emission cannot
be entirely from the background source.  Similarly, the high
resolution and high signal-to-noise submm images may provide a perfect
test-bed for lens mass profiles including substructure.  Nevertheless,
we leave more detailed lens modelling for future study.

\section{Results}
\label{sec_results}

\subsection{Lens model}

Table \ref{tab_lens_pars} lists the most probable lens model
parameters identified by marginalising over the Bayesian evidence
\citep[see][for more details]{nightingale15}. The parameters are
consistent with those obtained by D14 from modelling solely
the HST data. They are also in agreement with the parameters obtained by
\citet{rybak15} who performed $uv$ modelling of the same ALMA data we have
analysed in the present work. This provides further strength to our
argument concerning the viability of lens modelling in the image plane
in this case.

\begin{table}
%\sffamily
\centering
\small
\begin{tabular}{cl}
\hline
Lens parameter & Value \\
\hline
$\kappa_0$ & $(0.86\pm0.04)\times 10^{10}$M$_\odot$kpc$^{-2}$ \\
$\epsilon$ & $1.25\pm0.04$\\
$\theta$ & $13^\circ \pm 2^\circ$ \\
$\alpha$ & $2.01\pm0.05$\\
$\gamma$ & $0.04\pm0.01$\\
$\theta_\gamma$ & $-4^\circ \pm 3^\circ$\\
\hline
\end{tabular}
\caption{\small Most probable lens model parameters.  The quantities
  are the lens mass normalisation, $\kappa_0$, the elongation of the
  lens mass profile, $\epsilon$, the orientation of the semi-major
  axis of the lens measured counter-clockwise from north, $\theta$,
  the density profile slope, $\alpha$, the strength of the external
  shear component, $\gamma$ and the orientation of the shear
  $\theta_\gamma$ measured counter-clockwise from north.}
\label{tab_lens_pars}
\end{table}

We also note that the ring is remarkably well fit if we force a slope
of $\alpha=2$ and zero external shear, corresponding to a pure SIE
model. In this case, the best fit elongation increases to $\epsilon
\simeq 1.4$ but the model has a lower evidence (with $\Delta \chi^2
\simeq 20$) than the most probable model.

\begin{figure*}
\epsfxsize=17cm
{\hfill
\epsfbox{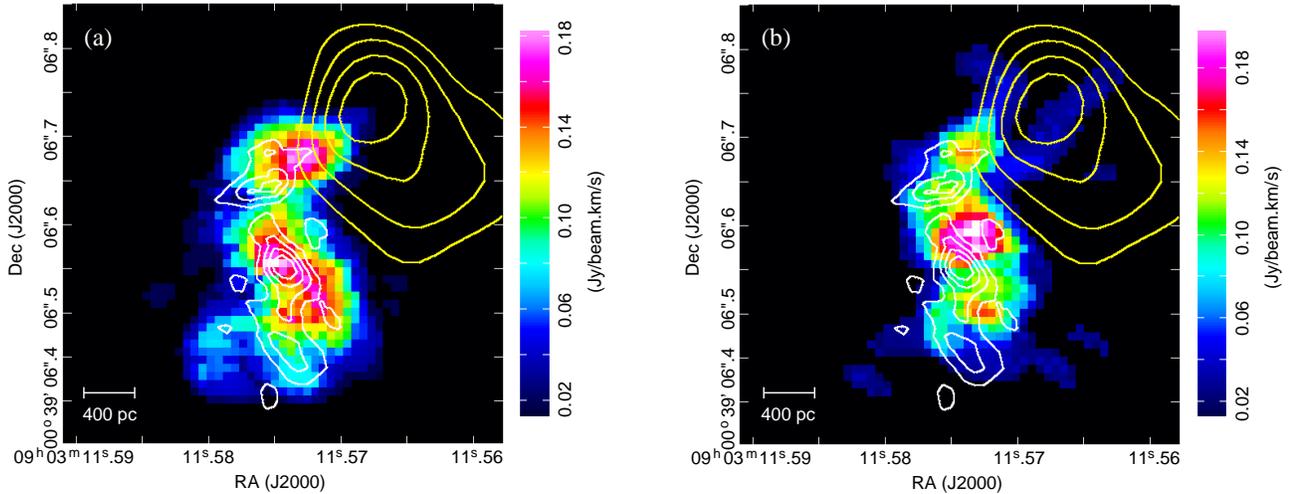}
\hfill}
\epsfverbosetrue
\caption{Zeroth moment CO maps for C0(5-4) (panel a) and CO(8-7)
  (panel b) emission in the source, clipped at $2\sigma$.  In both
  maps, the band 6 continuum emission is shown by the white contours
  and the F160W emission is shown by the yellow contours to illustrate
  the offset between the different wavelengths.}
\label{velocity}
\end{figure*}

\subsection{Source morphology}
\label{sec_src_morph}

Figure \ref{band6+7_cont_recon} shows the lensed source reconstructed
from the ALMA band 6 and band 7 continuum images. In the figure, we
show the source reconstructed on a regular grid for purposes of
illustration, although the adaptive source pixellisation scheme was
used in the lens modelling. We also show a zoomed-in section of
the image of the source to demonstrate the quality of the fit.

The source continuum emission shows that the distribution of dust is
very irregular with many small scale clumps. The morphology is broadly
similar between both bands, although the ratio of the band 7 flux to
the band 6 flux varies slightly across the source, most likely due to
varying dust temperature. \citet{rybak15} also show that the star
formation rate varies considerably across the source.

Turning to the F160W data, Figure \ref{hst_recon} shows our
reconstruction performed using the best fit lens model obtained by
simultaneously fitting to the band 6 and band 7 continuum data. As is
immediately apparent from a direct comparison of the observed lensed
images, there is a very distinct offset between the submm emission and
the near-IR (rest frame blue) emission. The additional structure
revealed by our re-processing of the F160W data is reconstructed in
the source plane as shown in panels (b) and (c) of Figure
\ref{hst_recon}.  Panel (c) in the figure reveals the dominant near-IR
component which gives rise to the bulk of the light seen in the
ring. This matches that reconstructed by D14, although a tail of
emission to the south is now apparent. In the larger scale
reconstruction shown in panel (b), it appears that this tail is only
part of a larger extent of near-IR emission.  In addition, to the
north of the main source component, there is another single component
which, in the image plane, gives rise to the arclet seen to the north
of the ring. We have measured the F110W-F160W colour of these
additional components, drawing on the shallower F110W data described
in N14, and find that it is consistent with the colour of the main arc
seen in the HST data.

We find total magnification factors of $16.0\pm0.7$ and $15.8\pm0.7$
for the band 7 continuum and band 6 continuum reconstructed sources
respectively. For the F160W source, we find a total magnification
factor of $4.5\pm0.3$ for the entire source as shown in panel (b)
of Figure \ref{hst_recon}. For the dominant component seen in
the F160W source, shown in panel (c) of Figure
\ref{hst_recon}, we measure a magnification factor of $10.2\pm0.5$,
consistent with the magnification measured by D14 for this part of the
source.

\subsection{CO emission and source kinematics}

We used our best fit lens model to reconstruct the distribution of
flux in the source plane for each slice of the image data
cubes released as part of the ALMA science verification data. Our
reconstruction was able to detect significant CO(5-4) and CO(8-7)
emission in the reconstructed band 4 and 6 cubes respectively. We were
unable to detect any significant CO(10-9) emission in the
reconstructed band 7 cube.

Figure \ref{velocity} shows the zeroth moment map of the CO line
emission.  These were made using {\tt CASA immoments}, stacking over
channels 31 to 61 inclusive (rest-frame velocities from
-370\,kms$^{-1}$ to 260\,kms$^{-1}$) for CO(5-4) and channels 30 to 60
inclusive (rest-frame velocities from -391\,kms$^{-1}$ to
239\,kms$^{-1}$) for CO(8-7). The channel ranges were selected to
fully encompass the spectral range of detected CO emission from the
source.  Our modelling yields total magnification factors of
$13.7\pm0.6$ for the CO(5-4) flux and $13.1\pm0.6$ for the CO(8-7)
flux.

For comparison, in the zeroth moment maps we also show the the band 6
continuum reconstructed source (white contours) and the F160W source
(yellow contours). It is immediately apparent from these plots that
while the CO(5-4) emission follows the continuum emission, there is a
significant offset between the more highly excited CO(8-7) emission
and the continuum. Furthermore, the CO(8-7) map shows more extended
emission, linking the submm and near-IR emission regions.

Using {\tt CASA immoments} to generate a first moment map to obtain
the velocity field in each reconstructed CO cube, we also found a
relatively smooth variation in velocity across the source
(see Figure \ref{vel_models}). The velocity field in each cube
has the hallmarks of disk rotation and hence we fitted
rotation curves assuming rotation of a disk
(see Section \ref{sec_dyn_modelling}). Similar application of
{\tt CASA immoments} to generate the second moment map, shows a
velocity dispersion which peaks in the dynamical centre of the
source but also has strong peaks throughout.

\subsubsection{Dynamical modelling}
\label{sec_dyn_modelling}

To model the velocity field of the CO(8-7) and CO(5-4) and so estimate
the disk inclination and asymptotic rotational velocity, we fitted a
two dimensional model whose velocity field is described by a
combination of stars and gas such that $v^2\,=\,v^2_D\,+\,v^2_H$ where
the subscripts denote the disk and dark matter halo respectively (we
ignore any contribution of H{\sc i} to the rotational velocity).  For
the disk, we assumed that the surface density follows a Freeman
profile (Freeman 1970). For the halo, we assumed the \citet{berkert95}
density profile which incorporates a core of size $r_0$ and converges
to the \citet[][NFW]{nfw96} profile at large radii. For suitable
values of $r_0$, the Berkert profile can mimic the NFW or an
isothermal profile over the limited region of galaxy mapped by the
rotation curves.

This mass model has three free parameters: the disk mass, the core
radius of the dark matter halo and the central core density.  To fit the
two-dimensional velocity fields, we constructed a two-dimensional
kinematic model with these three parameters, but we also we allowed
the [$x$\,/\,$y$] centre of the disk, the position angle (PA) and the
disk inclination to be additional free parameters.  We constrained the
[$x$\,/\,$y$] dynamical centre to be within 0.5\,kpc of the band 7
continuum centroid and then fitted the two dimensional dynamics using
an MCMC code \citep[see][for further details]{swinbank12}

\begin{figure*}
\epsfxsize=17.6cm
{\hfill
\epsfbox{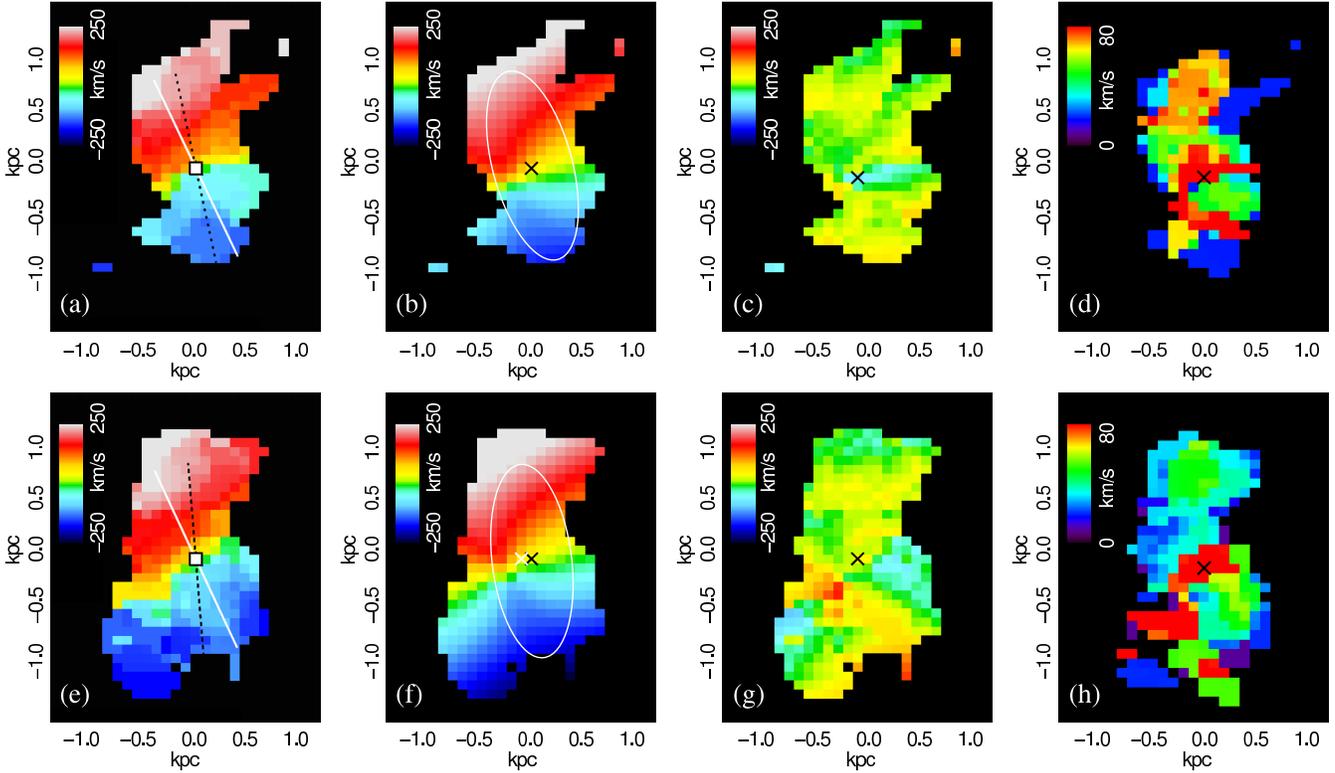}
\hfill}
\epsfverbosetrue
\caption{Observed, modelled and residual velocity fields (panels (a),
  (b) and (c) respectively) and observed line of sight velocity
  dispersion (panel d) for the CO(8-7) line emission.  The same
  quantities in the same order are shown for the CO(5-4) emission in
  panels (e) to (h).  The white line in panels (a) and (e) shows the
  principle axis of the best-fit disk model and the dashed black line
  shows that of the stacked CO emission. The white ellipse in panels
  (b) and (f) shows the extent of the disk used in the dynamical
  modelling.}
\label{vel_models}
\end{figure*}

The best-fit kinematic maps and velocity residuals are shown in Figure
\ref{vel_models}. The best fit disk inclination is
$(40\pm5)^\circ$. Assuming that the observed velocity field is indeed
due to disk rotation, then the observed maximum line of sight
rotational velocity of 210\,kms$^{-1}$ corresponds to an intrinsic
asymptotic velocity of 320\,kms$^{-1}$.

Figure \ref{rotn_curves} shows the one dimensional rotation curves
extracted from a 0.4\,kpc wide strip running along the major kinematic
axis identified from the dynamical models for the CO(8-7) and CO(5-4)
emission.  For these rotation curves, we defined the velocity zero
point using the dynamical centre of the galaxy. The error bars for the
velocities are derived from the formal $1\sigma$ uncertainty in the
velocity arising from the Gaussian profile fits to the CO emission.
We note that the data have not been folded about the zero velocity, so
that the degree of symmetry can be assessed.

\begin{figure}
\epsfxsize=7.5cm
{\hfill
\epsfbox{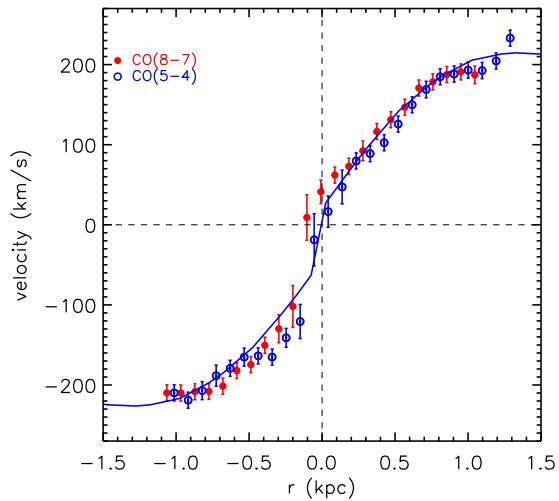}
\hfill}
\epsfverbosetrue
\caption{Rotation curves of the lensed source from CO(5-4) and CO(8-7)
  emission, strongly resembling disk-like dynamics in both cases. The
  data points show the observed line of sight velocity extracted from
  a 0.4\,kpc wide slit running along the major kinematic axis and the
  line shows the best dynamical model fit.}
\label{rotn_curves}
\end{figure}

The rotation curves imply a mass within 1.5\,kpc of $(3.5\pm0.5)\times
10^{10}$\,M$_\odot$ with only a small contribution from the halo within
this radius. Comparing the significantly lower maximum velocity
dispersion of $\sim 90$\,kms$^{-1}$ with this rotational velocity
suggests, under the assumption of a disk, that a correction for
asymmetric drift need not be applied to the inferred dynamical mass.
If we assume that all of the mass lies within the exponential disk
component, this corresponds to a surface mass density of
$5000\pm900$\,M$_\odot$pc$^{-2}$. This is a typical value for
high redshift SMGs \citep[see Figure 6 in][]{ivison13}.

Using this surface mass density and the observed velocity dispersion
in the disk, the resulting Toomre stability value is $Q \simeq 0.3$
which indicates a collapsing disk. Such low values of $Q$ are
observed in early merger systems before feedback can restore
equilibrium in the disk \citep[see][]{hopkins12}.

\subsection{Other intrinsic source properties}

\subsubsection{Dust mass and total dust luminosity}
\label{sec_dust_mass}

We have measured the dust mass of the lensed source by fitting a
two-component spectral energy distribution (SED) model to a
combination of our measured band 6 and band 7 continuum fluxes and
also the source fluxes presented in N14 but de-magnified by our average
continuum magnification factor of 15.9.  The SED model allows the dust
temperature, emissivity index, $\beta$, and optical depth at
100\,$\mu$m, $\tau_{100}$, to vary in the fitting. We used a dust mass
opacity coefficient equivalent to
$\kappa_{850\mu {\rm m}}=0.077$\,m$^2$\,kg$^{-1}$ to be consistent with the
fitting in N14.

\begin{figure}
\epsfxsize=8cm
{\hfill
\epsfbox{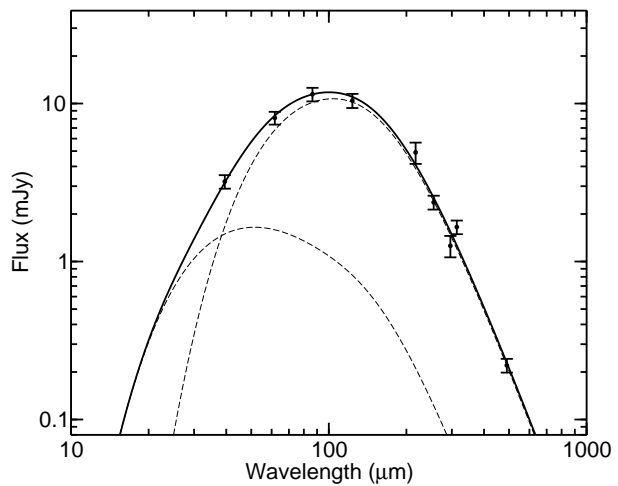}
\hfill}
\epsfverbosetrue
\caption{Rest-frame de-magnified best fit SED to the band 6 and 7
  continuum source fluxes and to the fluxes from table 2 in N14 (but
  de-magnified using our derived continuum source magnification
  factor). The dashed curves show the hot (99\,K) and cold (48\,K)
  SED components.}
\label{sed}
\end{figure}

Figure \ref{sed} shows the best fit SED model which has an effective
dust temperature of 51\,K, an emissivity index of $\beta=2.5$ and an
optical depth of $\tau_{100}=6.5$, yielding a dust mass of
$(1.8\pm0.3)\times 10^8$\,M$_\odot$ after de-magnifying by our average
continuum magnification factor. The fit favours both a hot and cold
dust component with temperatures of 99\,K and 48\,K respectively.

If we fix the emissivity index to $\beta=2.0$, we obtain a lower
optical depth of $\tau_{100}=3$ and a lower temperature of 46\,K but
all optically thick fits, with or without fitting to the 160$\mu$m
flux (which corresponds to rest-frame $\sim 40\,\mu$m and therefore
not well fit by a modified black body SED) returned a dust mass in the
range $(1.8-2.1) \times 10^8$\,M$_\odot$.  A second colder dust
temperature component (10--40\,K) was allowed in the fitting but was
not favoured by the data.

Assuming a typical gas to dust ratio of 150 \citep[for
  example][]{dunne00,draine07,coretese12,sandstrom13,swinbank14}
therefore gives a total molecular gas mass of $(2.7\pm0.5)\times
10^{10}$\,M$_\odot$. In making these calculations, we have of course
neglected any effects of differential magnification which could bias
the inferred dust temperature and/or emissivity index.

From integrating the best-fit SED between 8 and 1000\,$\mu$m, we
estimate that the total far-infrared emission is
$(5.0\pm0.8) \times 10^{13}$\,L$_{\odot}$, which agrees well
with the value of $5.4 \times 10^{13}$\,L$_{\odot}$ given
in N14. Our estimate of the average continuum magnification factor
of 15.9 implies that the intrinsic far-infrared luminosity
is $(3.1\pm0.5) \times 10^{12}$\,L$_{\odot}$.

\subsubsection{CO(1-0) gas}

We have derived an estimate of the CO(1-0) luminosity of the gas in
the lensed source using the CO(1-0) spectroscopic data acquired by
\citet{valtchanov11}. To do this, we firstly determined the spectrum
of the total lensed flux (note that Valtchanov et al. plot the peak
spectrum in their work). We then used our CO(5-4) source plane
velocity and the magnification map from the lens model to estimate the
magnification as a function of velocity in the source plane. By
transforming the total CO(1-0) SED to velocity space, we then
de-magnified the SED flux at each velocity with the corresponding
magnification factor from our magnification-velocity relation.  Figure
\ref{co1-0} shows the de-magnified SED.

Using the de-magnified SED, we determined a CO(1-0) flux of
$f_{\rm CO(1-0)} = 85 \pm 12$\,mJy\,kms$^{-1}$,
which corresponds to a luminosity of 
L$^{\prime}_{\rm CO}=(3.5\pm0.6)\times 10^{10}$\,K\,kms$^{-1}$\,pc$^2$.  
The errors take into account a number of uncertainties: 1) There
appears to be more CO(1-0) emission at higher and lower velocities
than we see in our CO(5-4) data and so in our velocity-magnification
relationship, we assumed a fixed magnification of 3 in these extremes;
2) The measure of de-magnified flux is sensitive to the binning of the
CO(1-0) spectrum and velocity-magnification mapping; 3) We have
determined the de-magnified flux assuming the CO(1-0) emission exactly
follows the CO(5-4) emission. The flux changes by $\sim 15\%$ if we
assume the CO(1-0) emission is more evenly distributed in the source
plane.

To estimate the total mass of molecular gas in the source, we used the
CO luminosity to molecular gas mass conversion factor from
\citet{bothwell13} of $\alpha_{\rm CO}=1$ \citep[although there is a
  very large uncertainty on this factor; see for
  example][]{papadopoulos12}. This gives a total molecular gas mass in
the source of $(3.5\pm 0.6) \times 10^{10}$\,M$_\odot$ which is larger
than the value of $(2.7\pm0.5)\times 10^{10}$\,M$_\odot$ we obtained
by scaling from the dust mass, although both come with considerable
uncertainty depending on the choice of assumptions (e.g., the
dust-to-gas ratio, or the minimum amplification for gas beyond that
seen in CO(5-4)).  Nevertheless, using the dynamical mass as an
estimate of the total mass, and the range of gas masses derived from
either the CO(1-0) or the dust luminosity, these values suggest a high
molecular gas fraction in the central regions, in the range $\sim
75-100$ per cent.  

\section{Discussion}
\label{sec_discussion}

%It has been known for almost two decades that high-redshift galaxies
%are highly clumpy and irregular in broad-band optical/UV images
%\citep{cowie95}. Despite this clumpiness, the ionised gas kinematics of
%these galaxies generally show that the galaxies are rotationally
%supported, although with much higher velocity dispersions than
%are seen in low-redshift galaxies \citep[see][and references
%  therein]{genzel14}. Observations in CO of star-forming galaxies at
%$1<z<2$, albeit with much lower resolution than the observations
%presented here, also reveal velocity fields as would be expected for
%rotationally supported disks \citep[see][and references
%  therein]{tacconi13}. Our analysis gives further support to the idea
%that most high-redshift galaxies are rotationally-supported disks.

\begin{figure}
\epsfxsize=8.5cm
{\hfill
\epsfbox{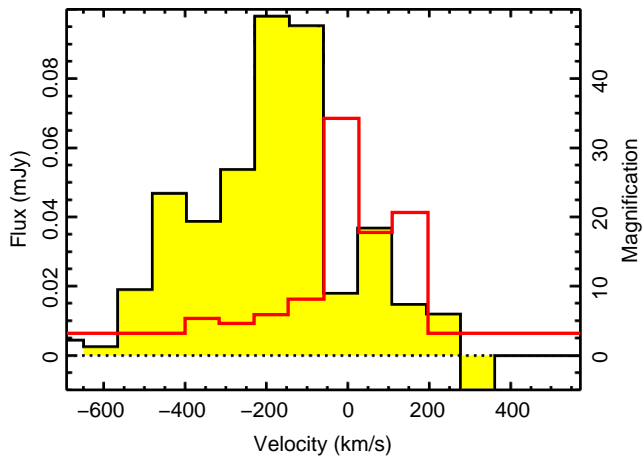}
\hfill}
\epsfverbosetrue
\caption{The de-magnified CO(1-0) spectrum in velocity space. The unshaded
red histogram shows our estimated magnification factor.}
\label{co1-0}
\end{figure}

The offset between the reconstructed rest-frame optical source
emission detected by HST/WFC3 and the reconstructed submm source from
the ALMA data is striking. D14 found good alignment between the
rest-frame optical emission and the submm source reconstructed by
\citet{bussmann13}, although this latter study used imaging acquired
with The Submillimeter Array (SMA) with a beam width approximately two
orders of magnitude larger than the ALMA image data analysed in the
present work. The anti-correlation seen between the rest-frame optical
and submm emission is not unique; the configuration in SDP.81 is
similar to the offset between the optical and submm emission seen in
the z=4.05 SMG GN20 \citep[see][]{daddi09, hodge15}. However,
occurrence of such large offsets is not common. For example, in a
survey of 126 SMGs carried out using ALMA \citep[see][]{hodge13}, only
two sources, ALESS88.11 and ALESS92.2 exhibit a similar configuration
\citep{chen15}.

There are a number of scenarios which could explain the observed
configuration of the lensed source. The three most likely ones are: 1)
The submm and optical components are simply two different sources
which are closely aligned on the sky but well separated along the line
of sight; 2) The submm and optical emission originates from the same
source and we observe a total lack of optical emission from the submm
region due to very strong dust absorption; 3) The optical and submm
components are two separate systems undergoing a merger.

The first scenario is challenging to verify, not least because of the
large effective radius of the lens in the optical/near-infrared
compared to the Einstein radius for the optical source. Photon noise
from the lens light thus dominates the already faint optical source
which will hinder attempts to measure its redshift, either
spectroscopically or photometrically. This scenario also draws into
question the structure of the optical source and the nature of the
apparent tidal debris to the south and the component to the north.

The second scenario could arise as a result of a strong dust lane in a
disk galaxy. There are many examples of ULIRGs in the local Universe
where strong absorption by dust lanes result in a delineation between
optical and submm emission. In the case of SDP.81, it may be that a
wide and thick dust lane is located at the eastern edge of the disk,
which, because of its inclination, provides a much higher column
density of dust toward regions in the source which emit optically. On
the assumption that the optical absorption efficiency factor of a dust
grain is inversely proportional to wavelength, which is true in Mie
theory if the grain size is much less than the wavelength, our
estimate of the optical depth at 100\,$\mu$m (see Section
\ref{sec_dust_mass}) implies that the optical depth in the optical
waveband is $\sim 1000$.  Therefore, we would not expect to see much
starlight from within the disk of gas and dust.

The third scenario is suggested by the presence of the northern
rest-frame optical component and the apparent tidal debris seen to the
south. A tempting interpretation of this is that the northern
component is a second galaxy which has passed through the larger,
strongly lensed system, causing the tidal debris observed and
enhancing the star formation rate and dust mass. Although the velocity
field appears to be quite regular, the low Toomre Q parameter we have
measured ($Q\simeq 0.3$) suggests a collapsing disk. Also, there are
several strong peaks in the velocity dispersion map which may still
point towards some level of interaction and there are filaments in the
CO(8-7) emission (but not the less excited CO(5-4) emission) which
extend up to the optical region. The strong emission in CO(8-7) is
itself consistent with a merger scenario since this higher energy
transition can not be produced over the entire observed disk by UV
flux from star formation \citep{papadopoulos14}.

In our attempts to search for more hints as to the nature of the
source, we also carefully searched through all channels in the band 6
and 7 ALMA data cubes, looking for CO emission from the northern and
southern optical components. Such emission might indicate an
association with the source or provide additional kinematic
information. We could not find any significant emission from 
the northern or southern optical components in either of the cubes.

Whatever the connection between the rotating disk of gas and dust
revealed by ALMA and the HST near-infrared sources, we can draw some
conclusions about the properties of the star formation in the disk. We
have estimated the star-formation rate in the disk from the intrinsic
far-infrared luminosity, which is the method suggested by
\citet{kennicutt12} for estimating the star-formation rate in a highly
obscured galaxy.  From the relationship between star-formation rate
and far-infrared luminosity given in Kennicutt \& Evans, we estimate
that the star-formation rate in this object is 
$\sim 470\pm80$\,M$_{\odot}$/yr.

\begin{table*}
%\sffamily
\centering
\small
\begin{tabular}{ll}
\hline
Source property & Value \\
\hline
Total continuum magnification & $15.9\pm0.7$ \\
Total rest-frame optical magnification & $10.2\pm0.5$\\
Total CO(5-4) magnification & $13.7\pm0.6$\\
Total CO(8-7) magnification & $13.1\pm0.6$\\
Disk inclination & $(40\pm5)^\circ$\\
Asymptotic rotational velocity & 320\,kms$^{-1}$\\
Toomre Q Parameter & 0.3\\
Total dust mass & $(1.8\pm0.3)\times 10^8$\,M$_\odot$\\
Total molecular gas mass from dust & $(2.7\pm0.5)\times 10^{10}$\,M$_\odot$\\
Dynamical mass within 1.5\,kpc & $(3.5\pm0.5)\times 10^{10}$\,M$_\odot$ \\
Total CO luminosity & L$^{\prime}_{\rm CO}=(3.5\pm0.6)\times
                          10^{10}$\,K\,km\,s$^{-1}$\,pc$^2$ \\
Total molecular gas mass from CO & $(3.5\pm 0.6) \times
                          10^{10}$\,M$_\odot$\\
Star-formation rate & $470\pm80$\,M$_{\odot}$/yr \\
CO(1-0) flux & $34\pm5$\,mJy\,kms$^{-1}$\\
Total far-infrared luminosity & $(3.1\pm0.5) \times 10^{12}$\,L$_{\odot}$\\
\hline
\end{tabular}
\caption{\small Summary of the intrinsic (i.e. de-magnified where 
relevant) physical properties of the source.}
\label{tab_source_props}
\end{table*}

Although it is well-known that high-redshift galaxies are very clumpy
and irregular in broad-band optical/UV images \citep{cowie95}, it has
always been an open question whether the clumpiness is the result of
the star formation occurring in clumps or whether it is the result of
patchy dust obscuration. In the case of this object, both the
reconstructed CO and dust emission are clumpy on the scale of the
point spread function in the reconstructed images, $\sim 200$\,pc. The
CO lines are high-excitation lines, so that we cannot rule out the
possibility that the clumpiness is the result of a variation in the
excitation rather than a variation in the distribution of the
gas. There is also the recent suggestion that a clumpy CO distribution
in high-redshift galaxies might be the result of the destruction of CO
molecules by cosmic rays \citep{bisbas15}. However, neither of these
two caveats apply to the dust emission; dust grains are robust and
found in all phases of the interstellar medium, and the emission from
the dust depends only very weakly on the intensity of the interstellar
radiation field. Therefore, the clumpiness of the dust is strong
evidence that the distribution of gas in this object is truly
extremely clumpy.  The low value of the Toomre Q parameter and the
very irregular distribution of gas are exactly what one would expect
if sections of the disk are collapsing to form stars. From the
rotation curve and the velocity dispersion, we estimate that the disk
is unstable over the scale range $\sim 50$\,pc to $\sim 700$\,pc, the
lower limit being the Jeans length and the upper limit being the scale
on which the disk should be stabilised by shear. This agrees well with
the sizes of the clumps observed.

Finally, we have estimated the efficiency of the star-formation
process in this galaxy. Using the dust mass as a tracer of the total
mass of gas, \citet{rowlands14} and \citet{santini14} have found
evidence that the star-formation efficiency (star-formation rate/gas
mass) is higher in high-redshift galaxies than in galaxies in the
local Universe. \citet{rowlands14} give relationships between the
star-formation rate and dust mass for local galaxies and for
high-redshift SMGs. Using these relationships, we estimate that a
typical galaxy in the local Universe with a dust mass equal to that of
SDP.81 would be forming stars at a rate of $7.2\pm1.3$\,M$_{\odot}$/yr
and that a typical SMG with the same dust mass would be forming stars
at a rate of $58\pm10$\,M$_{\odot}$/year. Thus our estimated SFR of
SDP.81 of 470\,M$_{\odot}$/yr implies a star formation efficiency that
is $\sim 8$ times greater than a typical SMG and $\sim 65$ times
greater than in the nearby Universe.

\section{Summary}
\label{sec_summary}

We have used the exceptional angular resolution of ALMA to reconstruct
a detailed map of the submm emission and dynamics in the lensed source
in SDP.81. Our modelling of the reprocessed HST data has revealed an
offset of $\sim 1.5$\,kpc between the submm and rest-frame optical
centroids in the source. The submm continuum emission in the source is
magnified by a total magnification factor of $15.9\pm0.7$ which
compares to the magnification of the rest-frame optical emission of
$10.2\pm0.5$ which mainly lies outside of the source plane
caustic. Similarly, the CO(5-4) and CO(8-7) emission is magnified by
the total magnification factors $13.7\pm0.6$ and $13.1\pm0.6$
respectively.

Our reconstruction of the source kinematics from the CO emission
reveals a relatively smooth velocity gradient across the source and
suggests regular disk-like rotation.  We have carried out dynamical
modelling of the observed line of sight velocities and find that the
data are best fit by a disk inclined at an angle of $(40\pm5)^\circ$
to the line of sight with an asymptotic rotational line of sight
velocity of 210\,kms$^{-1}$. Accounting for the disk inclination, this
corresponds to an intrinsic asymptotic velocity of 320\,kms$^{-1}$ and
an implied dynamical mass of $(3.5\pm0.5)\times 10^{10}$\,M$_\odot$
within 1.5\,kpc. Our dynamical modelling returns a low Toomre
Q-parameter of $Q\simeq 0.3$.

We have combined our measurements of the dust continuum flux from the
ALMA data with photometry of the lensed source given in
\citet{negrello14} to fit a modified black body SED. This indicates a
total dust mass of $(1.8\pm0.3)\times 10^8$\,M$_\odot$ after de-magnifying by
our average continuum magnification factor. Assuming a typical gas to
dust ratio of 150 gives total molecular gas mass of $(2.7\pm0.5)\times
10^{10}$\,M$_\odot$. We have also estimated the total molecular gas
mass from the de-magnified CO(1-0) spectrum of the lensed source from
\citet{valtchanov11}. This gives a total CO luminosity of
L$^{\prime}_{\rm CO}=(3.5\pm0.6)\times
10^{10}$\,K\,km\,s$^{-1}$\,pc$^2$ which, assuming a gas mass
conversion factor of unity, typical for ULIRGs in the local Universe,
gives a total molecular gas mass of $(3.5\pm 0.6) \times
10^{10}$\,M$_\odot$.

One observable we have not discussed in this work is the
low-excitation water line \watertrans\/ ($v_{\rm rest}$ = 987.927\,GHz
($E_{\rm up}$ = 101 K)) which can be seen in the band 6 data. We have
attempted to reconstruct the distribution of this line in the source
plane using our most probable lens model, but this has proven too weak
to locate easily. We have therefore left analysis of this feature for
future study.

To summarise, the nature of SDP.81 is somewhat perplexing! Although
the observational evidence we have assembled in this paper is
suggestive of a galaxy merger, we cannot rule out other possibilities.
Pending further analysis of additional observational data, the source
evades our full understanding. Nevertheless, this work has
demonstrated the complexity we can begin to expect in high redshift
SMGs when they are studied at the high angular resolution now made
possible by ALMA's incredible long baseline imaging capability.

\section*{Acknowledgements}

SD acknowledges financial support from the Midland Physics Alliance
and STFC. CF acknowledges funding from CAPES (proc. 12203-1).  MN
acknowledges financial support by PRIN-INAF 2012 project ‘Looking into
the dust-obscured phase of galaxy formation through cosmic zoom lenses
in the H-ATLAS'. LD, RJI and IO acknowledge support from the European
Research Council (ERC) in the form of the Advanced Investigator
Program, COSMICISM.  IRS acknowledges support from STFC
(ST/L00075X/1), the ERC Advanced Investigator programme DUSTYGAL
321334 and a Royal Society/Wolfson Merit Award. This paper makes use
of the following ALMA data: ADS/JAO.ALMA\#2011.0.00016.SV. ALMA is a
partnership of ESO (representing its member states), NSF (USA) and
NINS (Japan), together with NRC (Canada), NSC and ASIAA (Taiwan), and
KASI (Republic of Korea), in cooperation with the Republic of
Chile. The Joint ALMA Observatory is operated by ESO, AUI/NRAO and
NAOJ.  The work in this paper is based on observations made with the
NASA/ESA Hubble Space Telescope under the HST programme \#12194.

\label{lastpage}

\end{document}